\documentclass[a4paper,fleqn,usenatbib,useAMS]{mnras}

\usepackage{amssymb}	
\usepackage{multicol}        
\usepackage{bm}		
\usepackage{pdflscape}	
\usepackage[T1]{fontenc}
\usepackage{ae,aecompl}

\usepackage {graphicx}
\usepackage{natbib}
\usepackage{aas_journals}
\usepackage{color}
\usepackage{amsmath}
\newcommand{\OM}{\Omega_{\rm M}}
\newcommand{\OL}{\Omega_\Lambda}

\def\aap{{ A\&A}}

\def\apjl{{ApJL}}

\def\apjs{{ApJS}}

\newcommand{\bes}{\begin{equation*}}
\newcommand{\ees}{\end{equation*}}
\newcommand{\bea}{\begin{eqnarray}}
\newcommand{\eea}{\end{eqnarray}}
\newcommand{\beas}{\begin{eqnarray*}}
\newcommand{\eeas}{\end{eqnarray*}}

\newcommand{\ltsima}{$\; \buildrel < \over \sim \;$}
\newcommand{\lsim}{\lower.5ex\hbox{\ltsima}}
\newcommand{\gtsima}{$\; \buildrel > \over \sim \;$}
\newcommand{\gsim}{\lower.5ex\hbox{\gtsima}}

\def\gtrsim{\mathrel{\hbox{\rlap{\hbox{\lower4pt\hbox{$\sim$}}}\hbox{$>$}}}}
\let\ga=\gtrsim
\def\lesssim{\mathrel{\hbox{\rlap{\hbox{\lower4pt\hbox{$\sim$}}}\hbox{$<$}}}}
\let\la=\lesssim

\definecolor{mygray}{gray}{0.5}

\newcommand{\al}{\alpha}
\newcommand{\per}{\perp}

\newcommand{\be}{\begin{equation}}
\newcommand{\ee}{\end{equation}}
\newcommand{\ba}{\begin{eqnarray}}
\newcommand{\ea}{\end{eqnarray}}

\title[Cosmic troublemakers]{Cosmic troublemakers: the Cold Spot, the Eridanus Supervoid, and the Great Walls}

\author[Andr\'as Kov\'acs \& Juan Garc\'ia-Bellido]{Andr\'as Kov\'acs$^{1}$\thanks{E-mail: akovacs@ifae.es}, Juan Garc\'ia-Bellido$^{2}$\thanks{E-mail: juan.garciabellido@uam.es},\\
$^{1}$ Institut de F\'isica d'Altes Energies, The Barcelona Institute of Science and Technology, E-08193 Bellaterra (Barcelona), Spain\\
$^{2}$ Instituto de F\'isica Te\'orica IFT-UAM/CSIC, Universidad Aut\'onoma de Madrid, Cantoblanco 28049 Madrid, Spain
}

\begin{document}

\date{Submitted 2015}

\pagerange{\pageref{firstpage}--\pageref{lastpage}} \pubyear{2015}

\maketitle

\label{firstpage}
\begin{abstract}

The alignment of the CMB Cold Spot and the Eridanus supervoid suggests a physical connection between these two relatively rare objects. We use galaxy cata\-logues with photometric (2MPZ) and spectroscopic (6dF) redshift measurements, supplemented by low-redshift compilations of cosmic voids, in order to improve the 3D mapping of the matter density in the Eridanus constellation. We find evidence for a supervoid with a significant elongation in the line-of-sight, effectively spanning the total redshift range $z<0.3$. Our tomographic imaging reveals important substructure in the Eridanus supervoid, with a potential interpretation of a long, fully connected system of voids. We improve the analysis by extending the line-of-sight measurements into the antipodal direction that interestingly crosses the Northern Local Supervoid at the lowest redshifts. Then it intersects very rich superclusters like Hercules and Corona Borealis, in the region of the Coma and Sloan Great Walls, as a possible compensation for the large-scale matter deficit of Eridanus. We find that large-scale structure measurements are consistent with a central matter underdensity $\delta_0 \approx -0.25$, projected transverse radius $r_{0}^{\perp}\approx 195$ Mpc/h with an extra deepening in the centre, and line-of-sight radius $r_{0}^{\parallel}\approx500$ Mpc/h, i.e. an ellipsoidal supervoid. The expected integrated Sachs-Wolfe imprint of such an elongated supervoid is at the $\Delta T_{\rm ISW} \approx -40 \mu K$ level, thus inappropriate to accounting for the Cold Spot pattern in the CMB.

\end{abstract}

\section{Introduction}

The almost perfect cosmos reported by the Cosmic Microwave Background (CMB) analyses of the {\it Planck} mission comprises potential departures from isotropic and/or Gaussian statistics. These ``anomalies'' typically correspond to large angular scales, and their significance is subject to active research \citep[and references therein]{Schwarz2015}. One of these ripples is the Cold Spot (CS) centred on $(l,b) \simeq (209^\circ,-57^\circ)$ Galactic coordinates. It was first detected in the Wilkinson Microwave Anisotropy Probe \citep{bennett2012} maps at $\simeq 2-3\sigma$ significance using wavelet filtering \citep{VielvaEtal2003, CruzEtal2004}, and its presence was later confirmed by {\it Planck} \citep{Planck23}. However, \cite{ZhangHuterer2010} and \cite{Nadathur2014} have found that the most anomalous nature of the CS is in fact not its coldness at its centre, but rather the hot ring feature that surrounds it.

Proposals for the physical origin of the CS include rather exotic physics, e.g., textures \citep{CruzEtal2008,Vielva2010}, and combinations of the linear Integrated Sachs-Wolfe (ISW) \citep{SachsWolfe} and non-linear Rees-Sciama (RS) effects \citep{ReesSciama} from a deep $\gtrsim 200$ Mpc/h supervoid centred on the CS \citep{InoueSilk2006,InoueSilk2007,InoueEtal2010}, as a possible manifestation of Dark Energy's dominance at low redshifts. Such a deep void has not been found in large-scale structure analyses of the area, but there is firm evidence for a large but shallow supervoid in constellation Eridanus \citep{SzapudiEtAl2014, FinelliEtal2014}. However, a low-redshift spherical supervoid of $\delta_{0}\approx -0.25$ and $r_{0}\approx 200$ Mpc/h is capable of producing a $\Delta T_{\rm ISW} \approx -20 \mu K$ imprint in the CMB with dissimilar angular profile assuming spherical shape and concordance $\Lambda$CDM cosmology \citep{FinelliEtal2014,Nadathur2014,Zibin2014, MarcosCaballero2015,SzapudiEtAl2014}. 

Although no supervoid has been found that could fully account for the CS decrement, there is independent statistical evidence that super-structures imprint on the CMB as cold and hot spots \citep{GranettEtal2008,Planck23,CaiEtal2013}. The imprinted temperature in all of these studies is significantly colder for supervoids than simple estimates from linear ISW would suggest, therefore potentially represent a unique chance to learn about Dark Energy \citep[e.g.,][]{RudnickEtal2007,PapaiSzapudi2010, PapaiEtal2011,Flender2013,Hotchkiss2015}. The presence of this apparent excess ISW-like signal from supervoids and superclusters have generated debate about its physical origin in favor of a rare statistical fluke \citep{Nadathur2012,HM2013,Aiola2015}. A possible way out of this situation is the consideration of elongated void models \citep{Flender2013} and studies of potential biases in how void finder algorithms perform when applied to photometric redshift data \citep{Kovacs2015}. 

In principle, even spherical voids can appear elongated in the line-of-sight with tracer catalogues of typical photo-z smearing at the $\sigma_{z}=0.05(1+z)$ level. Such smearing effects can also result in non-detections of typical voids in average or over-dense environments \citep{BremerEtal2010}. This effect might be reduced when considering LRG tracer catalogues with more accurate photometric redshifts. On the other hand, void finders appear to be more sensitive to systems of multiple voids lined up in our line-of-sight, or underdensities elongated in this preferred direction. \cite{Sanchez2016} demonstrated that systematic analyses of mock galaxy catalogues and new void finding techniques optimized for photo-z surveys data are very important future tools to effectively study gravitational lensing signal around voids. Possibly, these biases in the void identification, if understood, are in fact helpful to measure localized ISW imprints of large and elongated underdensities. In fact, \cite{Granett2015} estimated the ellipticity of the 50 \cite{GranettEtal2008} supervoids\footnote{http://ifa.hawaii.edu/cosmowave/supervoids/}, identified by the \texttt{ZOBOV}\footnote{http://skysrv.pha.jhu.edu/neyrinck/voboz/} \citep{ZOBOV} algorithm, and found an average line-of-sight (LOS, hereafter) elongation of $q=2.6\pm 0.4$. This finding might {\it partially} explain why the ISW-like signal in the direction of supervoids appears to be significantly colder than theoretical predictions. \cite{Flender2013} analyzed the ISW imprint of large voids found in N-body simulations, without modeling photometric redshift errors in details, and concluded that the assumption of sphericity may lead to an underestimate of the maximum possible signal. However, the enhanced ISW imprint of elongated voids was found to be insufficient to explain the measurement by \cite{GranettEtal2008}.

Similarly, \cite{MarcosCaballero2015} estimated the effects of ellipticity on the ISW pattern imprinted by the Eridanus supervoid, and found colder central imprints (linear dependence on the elongation) with unsatisfactory angular profiles that cannot reproduce the observed properties of the CS. However, it is in principle possible that the true extent of the Eridanus supervoid is significantly larger than previous estimates, and even colder ISW-RS contribution is expected. In this paper, we estimate the LOS size of the supervoid as well as the projected transverse extent, and argue that the shape of the supervoid might shed new light on this outstanding problem. 

On the observational ground, the extra deepening in the central $\theta<5^{\circ}$ galaxy counts in ref.~\cite{SzapudiEtAl2014} suggests that the void might be slightly deeper in its centre than the first estimate. Moreover, the lack of data at the lowest redshifts and the limited sky coverage of the PS1 photo-z catalogue at the CS resulted in significant uncertainties and possible biases in the estimation of the total extent of the Eridanus supervoid.

We demonstrate that photometric redshifts of better quality and available spectroscopic measurements at the lowest redshifts reveal important details about the angular and radial shape of the Eridanus supervoid, and its surroundings. We further analyze the neighborhood of the Eridanus supervoid by looking at the density field in the antipodal direction on the sky that is physically close to the low-z part of the supervoid. This continuation of the LOS that we use for analyzing the direction of the CS happens to traverse the region that is called the Northern Local Supervoid, indicating that the Eridanus supervoid and its gravitational potential has connections and influence on the antipodal part of the sky. We find that this antipodal region includes the very rich Corona Borealis supercluster (CBS) that is perhaps the most massive member of the Sloan Great Wall~\citep[SGW]{Gott2005}. The CBS is located in the intersection of the SGW and another chain of superclusters~\citep{Einasto2011} which includes e.g. the Hercules supercluster in the Coma Great Wall~\citep{Geller1989}. 

Note that the existence of the Sloan Great Wall was first found to be difficult to reconcile with the standard model \citep{Sheth2011} but this tension relaxed by studying N-body simulations \citep{Park2012}. Similarly, the Eridanus supervoid is a rare object in the local Universe~\citep{Sahlen2015}. 

A final piece of evidence for the potential connection between local supervoids and region of the Great Walls is the existence of the Draco supervoid in the Northern hemisphere, as noticed in \cite{FinelliEtal2014}. We show that this other large-angle underdensity in the volume traced by the WISE-2MASS survey is the neighboring structure of the Great Wall system from another side.

The paper is organized as follows. Modeling tools are defined in Section 2. Data sets and map making algorithms are described in Section 3; our observational results are presented in Section 4; the final section 5 contains a summary, discussion and interpretation of our results.

\begin{figure}
\begin{center}
\includegraphics[width=90mm]{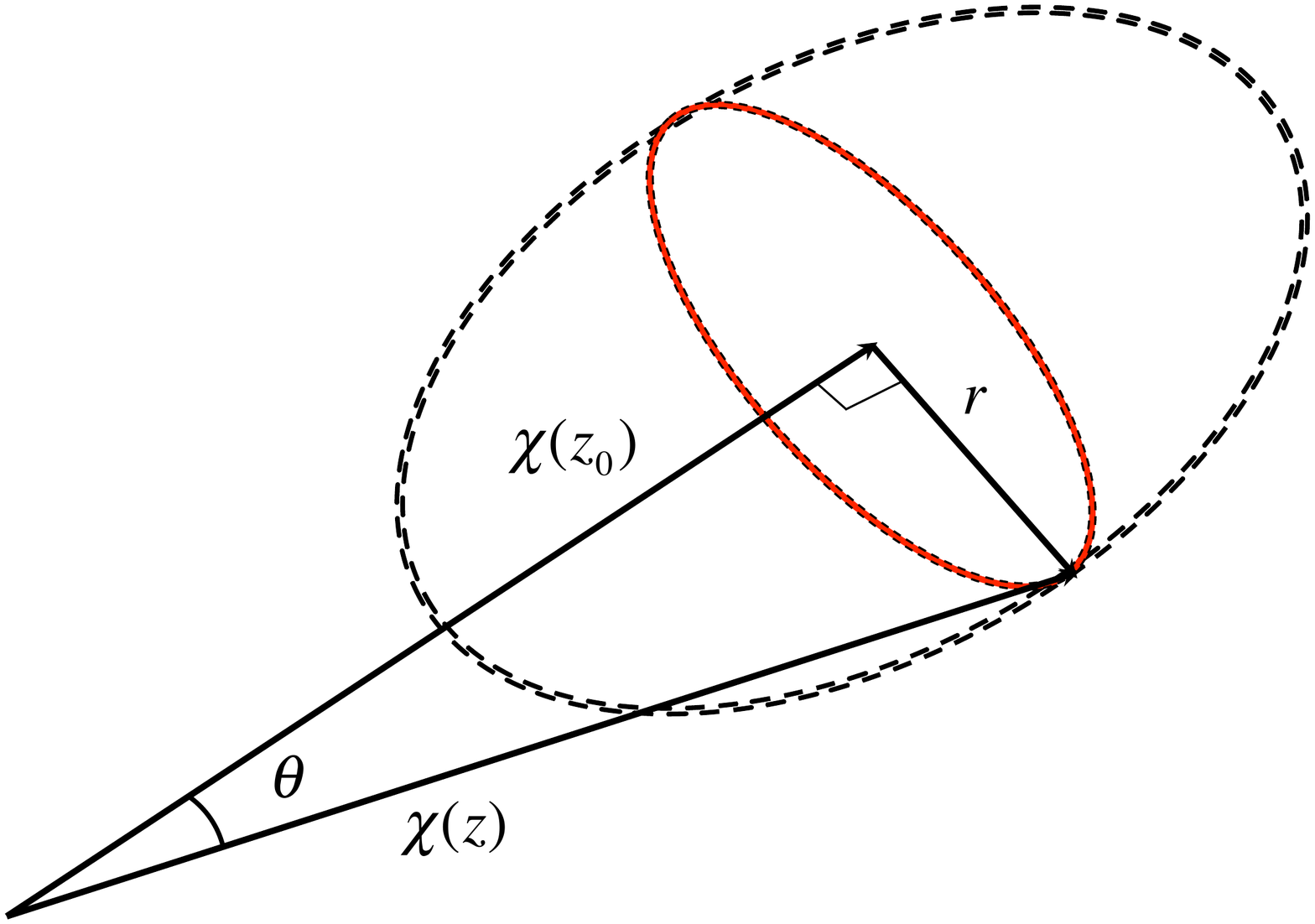}
\label{voidmodel}
\caption{Illustration of the ellipsoidal void model.}
\end{center}
\end{figure}

\section{Supervoid Model}
We follow \cite{FinelliEtal2014} in order to calculate the ISW and RS effects of supervoids, and model their profiles as a Gaussian profile with parameter $\al=0$,
\be
\Phi_0(r,z) = \Phi_0(z)\,\exp\Big[\!-\frac{r^2}{r_0^2}\Big]\,
\label{profilePhi_mod}
\ee
which corresponds to a density contrast:
\be
\delta(r) =  \delta_0\,g(a) \Big( 1 - \frac{2}{3}\,\frac{r^2}{r_0^2}\Big)\,\exp\Big[\!-\frac{r^2}{r_0^2}\Big]\,,
\label{profile3D_mod}
\ee
giving rise to a {\em compensated} void, with $g(a)$ the density contrast growth function of $\Lambda$CDM.

In general, we model a supervoid profile as an ellipsoidal Gaussian profile defined in Eq. (\ref{profilePhi_mod}).  Similar modeling was performed in \cite{MarcosCaballero2015}, also based mostly on \cite{FinelliEtal2014}. The ellipsoidal profile can be seen, in terms of the LOS and transverse coordinates (see also Fig.~1.),
\ba
&&r^2 = (1-e^2) r_\parallel^2 + r_\per^2 = \frac{1}{q^2} r_\parallel^2 + r_\per^2 \,, \label{r2mod}\\[1mm]
&&\hspace{3mm} r_\parallel = \chi(z) \cos\theta - \chi(z_0)\,, \\[1mm]
&&\hspace{3mm} r_\perp = \chi(z) \sin\theta \,, 
\ea
such that the radial coordinate in (\ref{profilePhi_mod}) and (\ref{profile3D_mod}) is given by
\be
\frac{r^2(z,\theta)}{r_0^2} =  \frac{(\chi(z)\cos\theta - \chi(z_0))^2}{q^2r_0^2} + \frac{\chi^2(z)\sin^2\theta}{r_0^2}\,,
\ee
with $\chi(z)$ the comoving coordinate in $\Lambda$CDM. The density contrast,
with these ellipsoidal modifications to the Gaussian profile (\ref{profilePhi_mod}), 
is given by the solution to the Poisson equation
\ba
\hspace{1cm} \delta(z,\theta)  \hspace*{-2mm} &\!=\!&  \hspace*{-2mm}
\delta_0\,g(a) \,\left( 1 - \frac{2}{1+2q^2}\,\Big(\frac{r_\parallel^2}{q^2r_0^2}+
\frac{q^2r_\perp^2}{r_0^2}\Big) \right) \times  \nonumber\\[2mm]
&& \hspace{9mm} \exp\left[-\,\Big(\frac{r_\parallel^2}{q^2r_0^2}+
\frac{r_\perp^2}{r_0^2}\Big)\right]\,, \label{densprofile}
\ea
with
\be\label{Phi0q}
\Phi_0(\delta_0,r_0) = - \left(\frac{3q^2}{1+2q^2}\right)\frac{\OM}{4}\frac{\delta_0\,H_0^2r_0^2}{g(1)}\,g(a)\,.
\ee
We define the transverse and along the LOS radial sizes of the void as the values they take when the density contrast crosses zero,
\ba
r_0^\perp \hspace{-1mm} &\!\equiv\!&  \hspace{-1mm} r_0\,\left(\frac{1+2q^2}{2q^2}\right)^{1/2}  \simeq ~ r_0\,, \nonumber\\[2mm]
r_0^\parallel  \hspace{-1mm} &\!\equiv\!&  \hspace{-1mm} q\,r_0\,\left(\frac{1+2q^2}{2}\right)^{1/2} \simeq ~ q^2\, r_0\,,
\ea
for large ellipticities.

The density contrast along the LOS (for $\theta=0$) becomes
\ba
\hspace{1cm} \delta_{\rm rad}(z)  \hspace*{-2mm} &\!=\!&  \hspace*{-2mm}
\delta_0\,g(a) \,\left( 1 - \frac{2}{1+2q^2}\,\frac{(\chi(z) - \chi(z_0))^2}{q^2r_0^2} \right) \times  \nonumber\\[2mm]
&& \hspace{9mm} \exp\left[-\,\frac{(\chi(z) - \chi(z_0))^2}{q^2r_0^2}\right]\,. \label{profile3D_rad}
\ea

On the other hand, the ISW angular temperature profiles for the ellipsoidal void, 
are given by
\ba
&&\hspace*{-7mm}\frac{\delta T}{T}^{\rm ISW}\hspace*{-4mm}(\theta) \simeq \frac{3\sqrt\pi}{22}
\frac{H(z_0)\,\OL\ F_4(-\OL/\OM(1+ z_0)^3)}{H_0(1+ z_0)^4\ 
F_1(-\OL/\OM)} \frac{3q^3}{1+2q^2} \times  \nonumber  \\[2mm]
&&\hspace*{-7mm} 
\Big(1+{\rm erf}\Big[\frac{z_0}{H(z_0)\,q\,r_0}\Big]\Big)\delta_0 (H_0r_0)^3
\exp\Big[\!-\frac{r^2(z_0)}{r^2_0}\sin^2x\Big]\,, \label{dTISW}
\ea
in the small angle approximation, $\cos x \approx 1 - \frac{1}{2}x^2 + \frac{1}{24}x^4$, where $x=\pi\theta/180 \ll 2\sqrt3$, valid for $\theta < 180^\circ/\pi$. We have checked that for $q\sim1$, the difference with respect to the exact integral is less than $\sim 5$\% up to $\theta \sim 60^\circ$.

Note that the RS contribution is always subdominant compared to the ISW signal. We therefore only consider the ISW predictions for drawing our conclusions when comparing our measurements to theory.

The main limitation of this modeling framework is the isolated nature of underdensities surrounded by a mean density Universe with no other structures. While a significant fraction of voids shows radial density profiles without significant compensation, especially the large ones \citep[see e.g.][]{Nadathur2016}, the understanding of the ISW imprint of individual systems of voids and nearby over-densities is very limited, especially in such a complicated case that the Eridanus supervoid represents with all its sub-voids and intervening filamentary structures. 

As a future test, advanced analytical supervoid models with more flexibility in the description of the compensation could be validated using the Jubilee LRG mock catalogue and the corresponding ISW map based on ray-tracing \citep{Jubilee,Hotchkiss2015}. Such comparisons may include tests of cosmic variance and analyses of the role of void environment in the ISW imprint.

\begin{figure}
\begin{center}
\includegraphics[width=85mm]{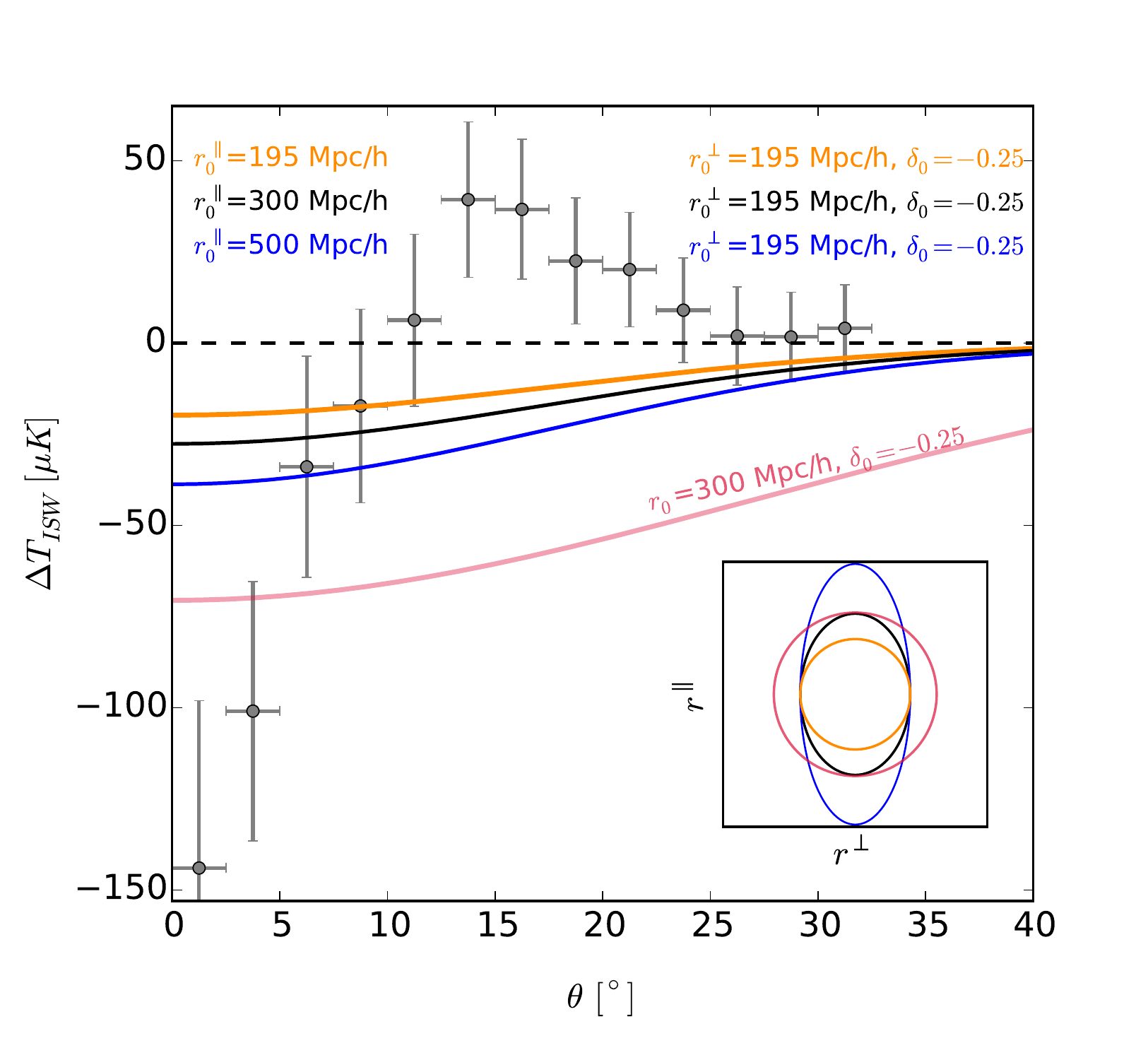}
\label{voidmodel}
\caption{Comparison of ISW imprint profiles of spherical and elongated model supervoids (solid lines) with the CS profile as measured in {\it Planck} data (grey points and error bars). The red curve shows the ISW expectation for a spherical supervoid of $r_{0}=300$ Mpc/h size, corresponding to the red circle in Figure~3 (the most extreme model considered by Marcos-Caballero et al.). Coloured text indicates the parameters of model supervoids, while the inset shows the actual relation of the models in terms of LOS vs. transverse size. Increased $r_{0}^{\parallel}$ with fix $\delta_{0}$ and $r_{0}^{\perp}$ results in larger ISW imprint based on Eq. (11).}
\end{center}
\end{figure}

\subsection{Parameter space for the model}

Checks for consistency with CMB measurements should fulfill three main criteria: 
\begin{enumerate}
\item the $\theta \approx5^{\circ}$ characteristic angular radius of the CS
\item the deep $\Delta T\approx -150 \mu K$ temperature depression in its centre
\item the $\Delta T\approx 30 \mu K$ hot ring feature that surrounds it at $\theta \approx15^{\circ}$
\end{enumerate}
Note that the spherical void model with $\delta_0 = -0.25$ and $r_{0}=195$ Mpc/h by \cite{FinelliEtal2014} is insufficient for explaining any of these features, as e.g. \cite{Nadathur2014} already pointed out. Nevertheless we fix the transverse size of the Eridanus supervoid to $r_{0}^{\perp}=195$ Mpc/h in our ellipsoidal modeling, based on the projected angular density profile of the supervoid that reaches the mean density at $\theta \approx25^{\circ}$ in WISE-2MASS data. We confirm this finding with 2MPZ data, and show the results in Figure~6. We also fix the central redshift to $z_{0}=0.14$.

In this modeling framework, a rather extreme spherical supervoid of $\delta_0 \approx -0.25$ and of radius $r_{0}\approx400$ Mpc/h leaves a central imprint as cold as the central CS temperature (see Figures~2 and 3). However, matching the angular extent of the CS is only possible with a rather narrow void, and there is no explanation for the CS hot-ring feature assuming these parameters in our model. See Figure~2 for details about the relation of different models assuming spherical shape or allowing for elongation in the LOS.

\begin{figure}
\begin{center}
\includegraphics[width=90mm]{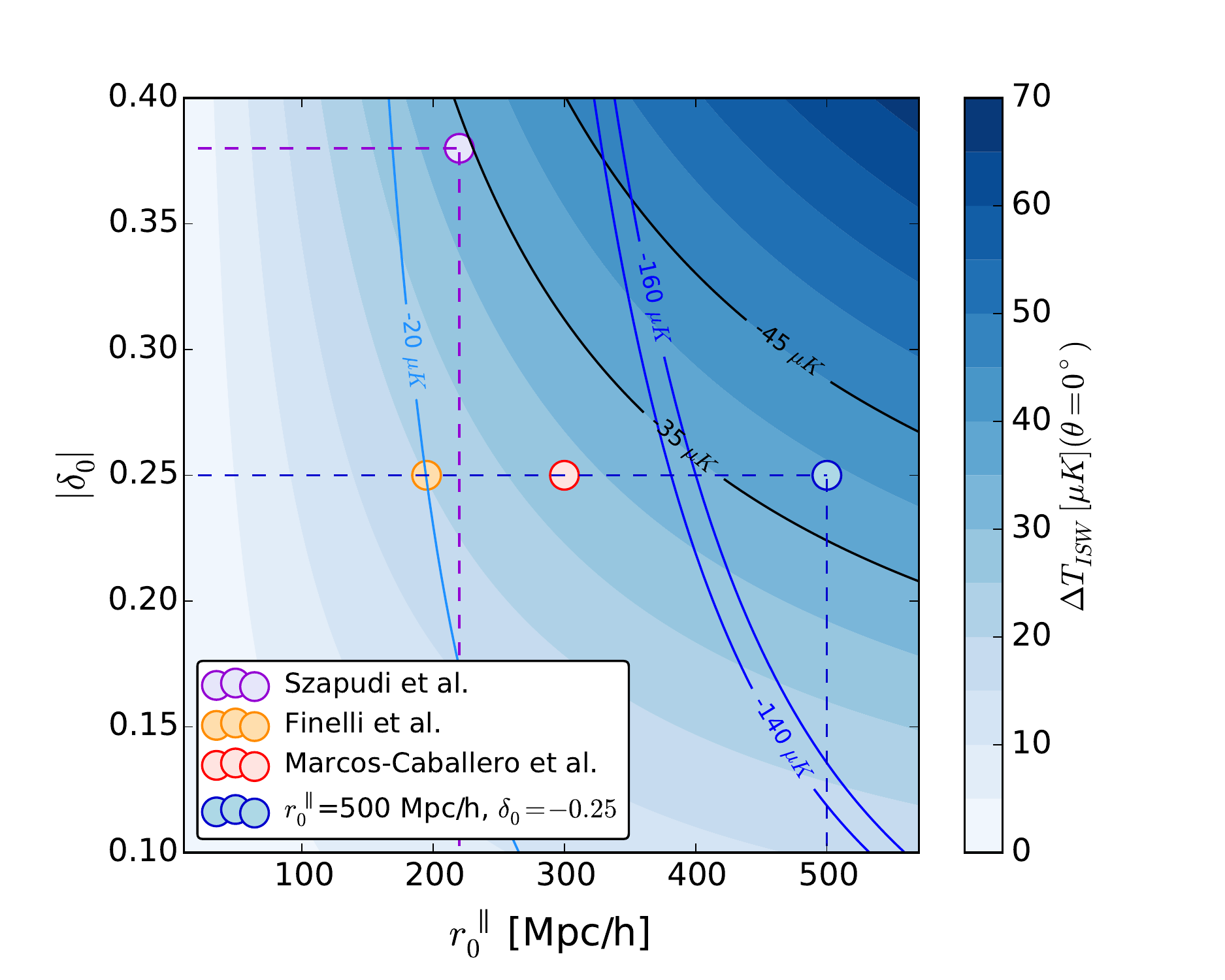}
\label{}
\caption{We show central ISW imprints of model supervoids at fixed central redshift $z_{0}=0.14$ and $r_{0}^{\perp}=195$ Mpc/h (numerical values are indicated by the colour bar). The parameters that we varied are the central underdensity and the LOS void radius. We mark the $-40 \pm 5~ \mu K$ range of the underlying colour map with the black contours. Blue contours correspond to spherical models with $r_{0}^{\perp}=r_{0}^{\parallel}$ to illustrate the important role of $r_{0}^{\perp}$ in the ISW signal (see also Fig. 2). Purple, red, and orange markers indicate previous models considered for predicting the ISW effect of the supervoid. For Marcos-Caballero et al., we show the most extreme radius that they consider. For Szapudi et al. we assumed $\delta_0 = -0.38$ and $r_{0} = 220$ Mpc/h.}
\end{center}
\end{figure}

In Figure~3, we present a comparison of central $\Delta T_{\rm ISW}$ imprints of {\it ellipsoidal} supervoids of fix $r_{0}^{\perp}=195$ Mpc/h transverse radius. Colour-coding indicates the central $\Delta T_{\rm ISW}$ signal of these objects. The $-40 \pm 5~ \mu K$ range of imprints, which is potentially relevant and achievable for supervoids like the $r_{0}^{\parallel}\approx500$ Mpc/h model we propose, is marked by the black contours.

We also estimate the imprints of {\it spherical} supervoids that have been proposed as explanations for the CS (light blue contour). Medium blue contours indicate the range $-150 \pm 10 \mu K$, where a hypothetical low-z supervoid, capable of explaining the CS central temperature, should be located. The imprint of ellipsoidal voids is significantly suppressed compared to their spherical counterparts with identical LOS radii. 

Note that the probability of spherical underdensities of this size is extremely low in the $\Lambda$CDM cosmological model \citep{Nadathur2014}. In principle, \cite{bardeen} provide a general framework to model ellipsoidal density fluctuations at linear scales, but very little is known about the observable shape of the largest negative density fluctuations, that surround the largest superclusters often with multi-spider morphology~\citep{Einasto2011}. Most studies have been focusing on the identification of smaller individual voids in the cosmic web, and not on the possible connections between void systems. A recent example for such approach was presented in \cite{Nadathur2016} by studying "Minimal" voids with specific merging criteria.

We seek large-scale structure evidence for such an elongated supervoid, or system of fully connected voids.

\section{Data sets}

\subsection{Planck}

For {\it Planck} CMB temperature data we use the Spectral Matching Independent Component Analysis (SMICA)\footnote{http://pla.esac.esa.int/pla/aio/planckProducts.html} \citep{CardosoEtal2008,Planck2015I}
map downgraded to \texttt{HEALPIX}\footnote{http://healpix.jpl.nasa.gov} \citep{healpix} resolution $N_{\rm side}=128$.
We use a definition for the centre of the CS from the latest {\it Planck} results \citep{Planck24,Planck2015XVI}. Based on the literature, we decided in advance to test for an underdensity at $\theta<5^\circ$ around this centre to match the approximate angular size of the CS \citep{VielvaEtal2003, CruzEtal2004, RudnickEtal2007,SzapudiEtAl2014}. We minimize the {\it a posteriori} bias by setting this value from CMB independently of the large-scale structure data we use, and this choice also simplifies the interpretation of our measurements in the Bayesian framework.

\subsection{Previous galaxy surveys at the Cold Spot}

\cite{RudnickEtal2007} have found the first evidences for an underdensity by studying a catalog of radio galaxies in the NRAO VLA Sky Survey (NVSS)\footnote{http://www.cv.nrao.edu/nvss/}. However, no redshift information was available for the supervoid candidate, and its significance has been disputed \citep{SmithHuterer2010}. Targeted pencil beam surveys \cite{GranettEtal2010} and \cite{BremerEtal2010} found no significant underdensity between redshifts of $0.5 < z < 0.9$, but their data are consistent with a void at $ z < 0.3$. In addition, the analysis of the 2-Micron All-Sky Survey Extended Source Catalog\footnote{http://www.ipac.caltech.edu/2mass/releases/allsky/} \citep[2MASS XSC]{jarrett2000} galaxy distribution by \cite{francis2010} showed a shallow underdensity of large angular size around the CS. \cite{rassat2013} confirmed the presence of this low-z void in the reconstructed 2MASS ISW maps. The ISW imprint is consistent with $\Delta T_{\rm ISW}\approx -50 \mu K$ at the CS (the most significant imprint in the observable sky), if all multipoles $2<\ell<30$ are taken into account for a joint 2MASS-NVSS ISW reconstruction \citep{Rassat2014}.

Recently, \cite{FinelliEtal2014} identified a similarly large underdensity in the WISE-2MASS galaxy catalogue by \cite{KovacsSzapudi2014} that combines measurements of two all-sky surveys in the infrared, the Wide-Field Infrared Survey Explorer\footnote{http://wise.ssl.berkeley.edu} \citep[WISE]{wise}) and the Point Source Catalog of the 2-Micron All-Sky Survey \citep[2MASS-PSC]{2mass}. \cite{SzapudiEtAl2014} matched the WISE-2MASS galaxy map to a 1,300 deg$^2$ area with the PV1.2 reprocessing of Pan-STARRS1 \citep[PS1]{ps1ref}, adding optical colours for each object. In the resulting catalog with photometric redshifts \cite{SzapudiEtAl2014} analyzed the LOS density profile in the redshift range $ z < 0.3$, and detected a supervoid of radius $r_{0}$=220 Mpc/h centred on the CS. 

Importantly, \cite{Manzotti2014} found that any late time ISW+RS imprints that might be responsible for the Cold Spot are very likely to be originated at $z < 0.3$, thus the most detailed examination of this region is requisite.

See Table~1 for a detailed comparison of current and previous galaxy samples used for mapping the CS region, including the Dark Energy Survey \citep[DES]{DES}. DES has started its observations in the region of interest, and will presumably provide important details about the Eridanus supervoid by mapping the density field traced by the redMaGiC selection of luminous red galaxies with accurate photometric redshifts \citep{Rozo2015}.

\begin{table}
\begin{center}
\begin{tabular}{lrrrrr}
\hline
Survey & $N_{gal}$ & CS map & $b_{g}$  & $z_{med}$ & LOS info \\
\hline
NVSS & $1.1\times 10^{6}$ & Full & $\sim 2$ & $\sim1.5$ & \O \\
Granett+ & $\sim 10^{4}$ & Partial & 1.5 & $\sim0.5$ & photo-z \\
Bremer+ & $7.3\times 10^{2}$ & Partial & -- & $\sim0.5$ & spec-z \\
2MASS &  $1.6\times 10^{6}$ & Full & 1.2 & 0.07 & photo-z \\
W2M & $2.4\times 10^{6}$ & Full & 1.4 & 0.15 & \O \\
W2M-PS1&  $7.3\times 10^{4}$ & Partial & 1.4 & 0.15 & photo-z \\
\hline
2MPZ &  $9.3\times 10^{5}$ & Full & 1.2 & 0.08 & photo-z \\
6dF &  $1.2\times 10^{5}$ & Full & 1.5 & 0.05 & spec-z \\
\hline
DES &  {$3\times 10^{8}$} & {Full} &{1.6} &{$\sim0.5$} & {photo-z} \\
\hline
\end{tabular}\caption{Properties of galaxy surveys that mapped the CS region. \label{table_surveys}}
\end{center}
\end{table}

\subsection{2MASS photometric redshift catalogue}
We extend the previous analyses of the CS region by using the 2MASS Photometric redshift catalogue\footnote{http://surveys.roe.ac.uk/ssa/ TWOMPZ} \citep[2MPZ]{Bilicki2014}, i.e. a low-redshift photo-z map for almost the full sky, powered by a WISE-2MASS-SuperCOSMOS matched catalog containing infrared colours $W1_{\rm wise}$, $W2_{\rm wise}$, $J_{\rm 2mass}$, $H_{\rm 2mass}$, $K_{\rm 2mass}$, and optical photometry $B_{\rm sc}$, $R_{\rm sc}$, and $I_{\rm sc}$~\citep{supercosmos}. The parent sample of 2MPZ is the 2MASS XSC that includes $\sim1.6\times 10^{6}$ resolved sources detected on most of the sky except the highly confused Galactic bulge. 

The 2MPZ catalog contains more accurate photometric redshifts than those of \cite{b14} who created a similar catalog without WISE observations. The accuracy reached in the photo-z estimation process with \texttt{ANNz}\footnote{http://www.homepages.ucl.ac.uk/~ucapola/annz.html} algorithms is $\sigma_{z}\approx 0.015(1+z)$, or $12\%$ on average. The redshift distribution of the 2MPZ catalog is shown in Fig.~4. The smoothed full sky distribution of the 2MPZ galaxies is presented in the middle panels Fig.~5.

We mask pixels with $E(B-V) \geq 0.1$ \citep{SchlegelEtal1998}, and regions at galactic latitudes $|b| < 10^{\circ}$ to exclude potentially contaminated regions near the Galactic plane, and the Magellanic Clouds. A stripe at $l=140^{\circ}$, $b \leq 35^{\circ}$ with a lack of WISE measurements due to torque rod gashes or incompleteness in WISE due to moon contamination was also masked out, following \cite{Bilicki2014}. 

For the linear bias parameter that we consider in our following analysis, we use the value of $b_{g}=1.18\pm0.03$ for $z<0.08$ and $b_{g}=1.52\pm0.03$ for $z>0.08$, as estimated by \cite{Alonso2015}. We note that a deterministic linear galaxy bias is found to be a poor assumption for deep voids \citep{Neyrinck2014,Nadathur2015}. Nevertheless, we assume such a relation throughout this paper because the Eridanus supervoid is rather shallow and we have no more information on its shape, density profile, and surroundings because of the measurement uncertainties.

\cite{Alonso2015} used the 2MPZ data for cosmological analyses and reported excellent agreement with $\Lambda$CDM predictions in terms of the galaxy auto-correlation function, dipole analysis of the local galaxy distribution, and homogeneity tests. We interpret these findings as confirmations of the high quality of the 2MPZ galaxy catalogue, including negligible systematic problems at the largest scales.

\begin{figure}
\begin{center}
\includegraphics[width=85mm]{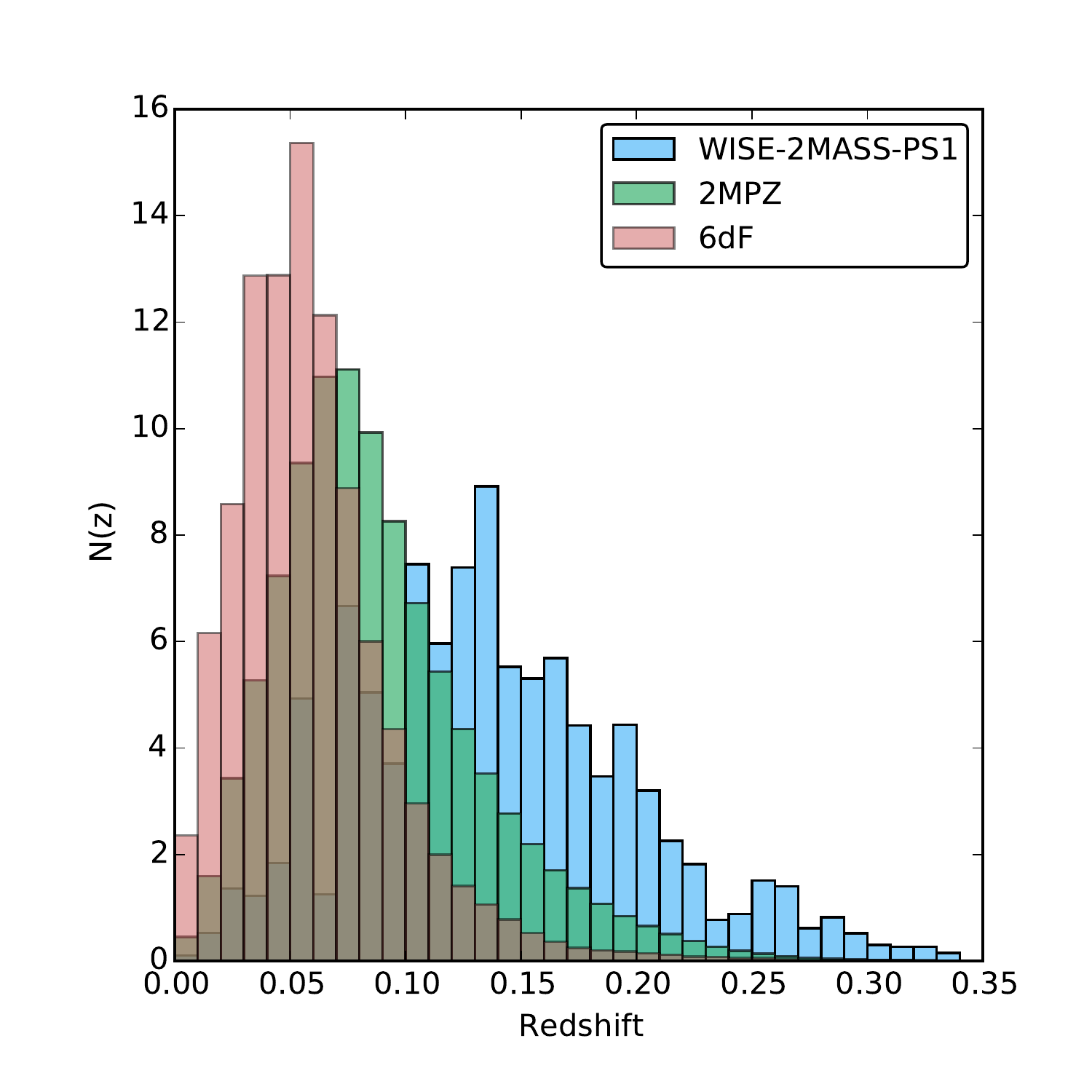}
\label{redshift}
\caption{Redshift distributions of galaxy samples tracing the Eridanus Supervoid are compared. We use the photo-z distribution of the 2MPZ galaxies, the spec-z sample of the 6dF survey, and the estimated redshift distribution of the WISE-2MASS catalogue.}
\end{center}
\end{figure}

\subsection{6dF Galaxy Survey}
For the mapping of the CS region at the lowest redshifts, we use the 6dF Galaxy Survey Data Release 3 (DR3) galaxy catalogue\footnote{http://www-wfau.roe.ac.uk/6dFGS/} \citep[6dFGS]{Jones2009} that is a near-infrared selected redshift survey covering 17,000 $deg^{2}$ in the Southern sky. The near-infrared photometric selection was based on the 2MASS XSC, while the spectroscopic redshifts of 6dFGS were obtained with the Six-Degree Field (6dF) multi-object spectrograph of the UK Schmidt Telescope (UKST). 

The survey avoids a $|b| \leq 10^{\circ}$ region around the Galactic plane to minimize extinction and foregrounds. For our density field analysis, we use the mask constructed for the 2MPZ sample supplemented by an extra cut at $Dec<0^{\circ}$ due to the 6dF survey boundary.

For this galaxy catalogue, we consider a deterministic linear galaxy bias parameter of $b_{g}=1.48\pm0.27$ as measured by \cite{Beutler2012}. In Fig.~2, we compare the redshift distribution of the 6dF catalogue to those of 2MPZ and WISE-2MASS-PS1. 

\begin{figure*}
\begin{center}
\includegraphics[width=180mm]{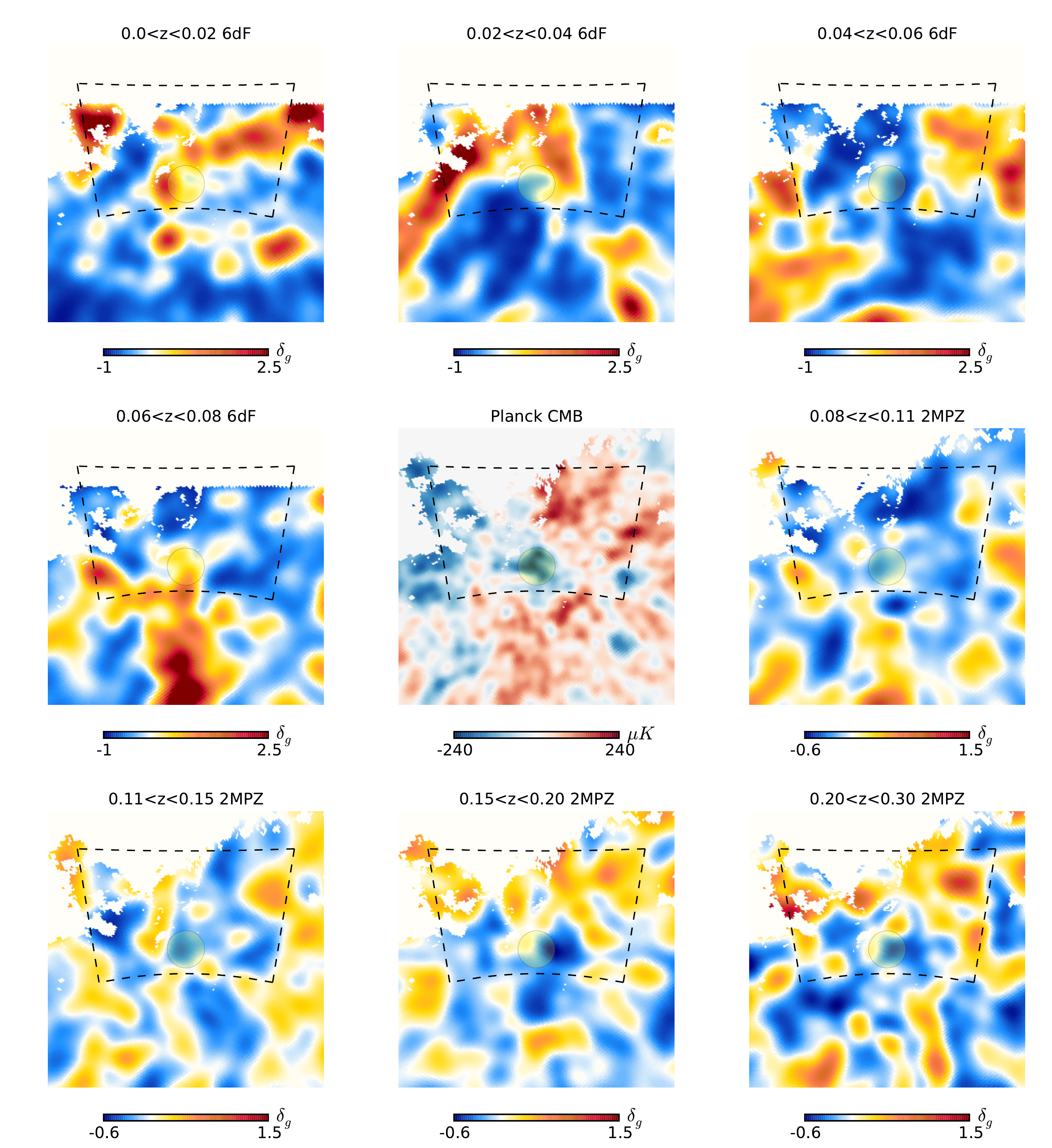}
\label{tomography}
\caption{Tomographic imaging of the wider CS region in boxes of size $75^{\circ}\times 75^{\circ}$ (in Equatorial coordinates) around the nominal centre at $(l,b)=(209^{\circ},-57^{\circ})$. We apply a Gaussian smoothing of $\sigma=2^{\circ}$ in order to show potential substructure in the galaxy density field. For redshifts $z<0.08$, we use 6dF galaxies, and 2MPZ density mapping otherwise. The yellow circle indicates the $\theta<5^{\circ}$ region around the centre. The black dashed area shows the approximate footprint of the WISE-2MASS-PS1 photo-z mapping by Szapudi et al. (2015), including the PS1 survey boundary at $Dec=-28^{\circ}$. The central image illustrates the CS region in the Planck CMB temperature map. In the main text, we add supplementary information about known prominent neighboring superstructures in each tomographic bin. The most significant intervening structure is one of the outer filamentary features rooted in the Horologium supercluster ($z\approx 0.07$).}
\end{center}
\end{figure*}

\begin{figure*}
\begin{center}
\includegraphics[width=190mm]{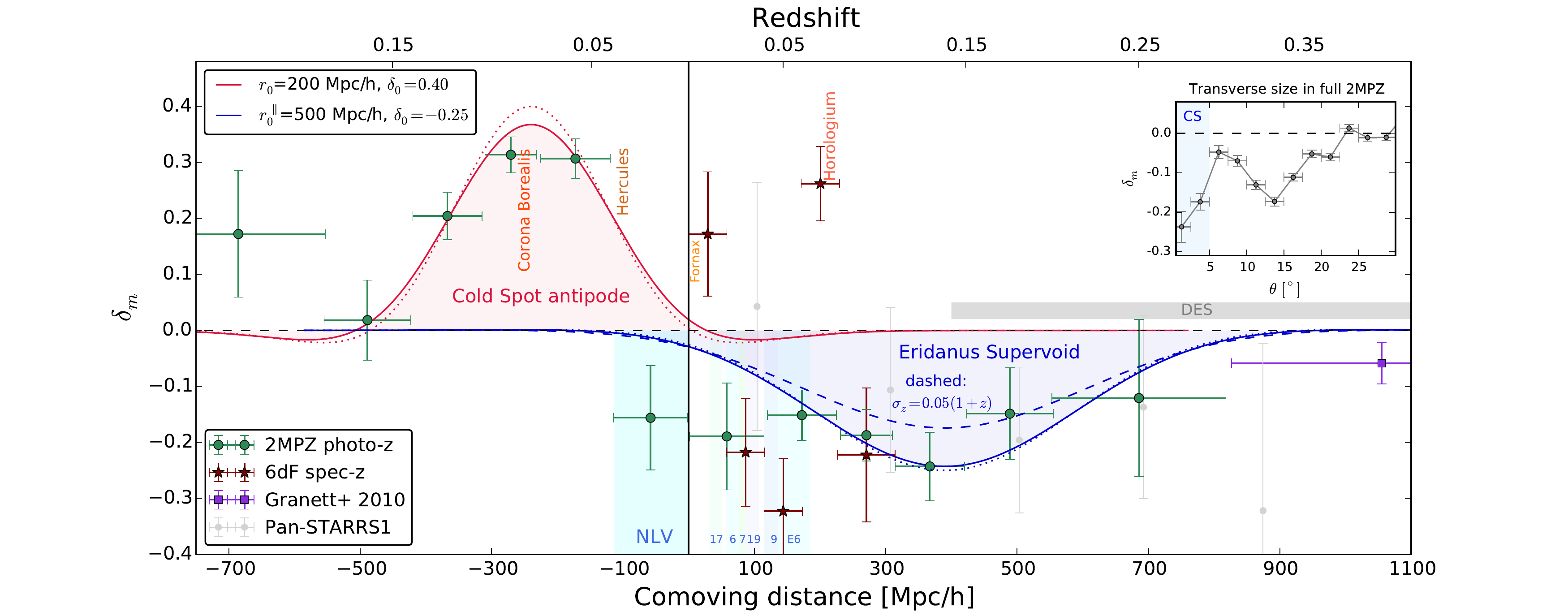}
\label{los_prof}
\caption{{\it Right:} Line-of-sight density profile measurements with Poisson error bars at the direction of the Cold Spot centre $l,b=209^{\circ},-57^{\circ}$ (right), and in the antipodal direction $l,b=29^{\circ},57^{\circ}$ (left). We compare results using various data sets. Shaded areas show the locations of significant nearby voids in the Eridanus constellation that were identified in EEDTA (code: $E6$) and SSRS2 (codes: $6,7,9,17,19$). Solid lines show our model profiles including photo-z smearing effects, while dotted lines mark the model without photo-z smearing, showing good agreement for such large structures. The dashed blue line shows the smearing effect of a hypothetical larger photo-z error on the model profile, now showing the limitation of photo-z analyses of much worse quality than in 2MPZ. The grey band marks out the redshift range where the Dark Energy Survey will most sensitively map further details in the LOS profile of the supervoid using LRGs. {\it Left:} Interestingly, the antipodal LOS begins by passing through the Northern Local (Super)void (NLV) region, i.e. a knowingly significant deficit of rich clusters in the local Universe (Lindner et al. 1994). More interestingly, the antipodal direction is partially aligned with the Hercules supercluster, the Corona Borealis supercluster, and therefore intersects the Sloan Great Wall. {\it Inset:} radial density profile in the {\it full projected} 2MPZ catalogue combining all redshift bins.}
\end{center}
\end{figure*}

\subsection{Super-structure catalogues}

We consider three catalogues of low redshift voids in order to complement the information in the galaxy catalogues with existing knowledge on local underdensities. Void positions, approximate size information, and their relation to nearby superclusters are really informative additional tools to confirm our findings below.

\begin{enumerate}
\item void catalogue by \cite{Batuski1985} ($B\&B$) that contains 29 voids defined by the absence of Abell clusters.
\item a similar compilation of cluster-defined voids appears in \cite{Einasto1994} (EEDTA), i.e. a census listing a total of 27 supervoids within a cube of 740 Mpc a side.
\item we consider voids in the Eridanus constellation defined by spec-z tracer galaxies of the Southern Sky Redshift Survey 2 (SSRS2) \citep{ElAd1997}.
\end{enumerate}

\section{Results}

\subsection{Tomography}

We first perform a tomographic imaging in the wider CS area of $75^{\circ}\times 75^{\circ}$ using 2MPZ and 6dF data. The resulting maps, spanning $z<0.3$ in eight redshift slices of different width, are shown in Fig.~5. We note the following features in this analysis:
\begin{itemize}
\item the LOS we analyze passes by the Eridanus group and the Fornax cluster in the lowest bin at $z<0.02$, but the wider CS region is underdense.
\item at $0.02< z <0.04$ the 6dF data shows that the CS LOS avoids other overdensities in the very nearby Universe.
\item the CS region slightly overlaps with one of the outer filamentary arms of the Horologium supercluster.
\item at $0.08< z < 0.11$, now traced by 2MPZ photo-zs, the Horologium supercluster is still visible in the bottom of the image, while the CS LOS traverses the centre of an extended system of underdensities with deeper subvoids.
\item the large-scale underdensity is followed by a set of typical realizations of over- and underdensities of smaller angular size, but the CS area is still underdense.
\item at $0.15< z <0.20$, a deep, extended, and slightly miscentred void is observable, surrounded by presumably prominent superclusters.
\item the last photo-z bin, where 2MPZ starts running out of data, again shows that the CS area passes through the centre of an extended underdensity.
\end{itemize}

The topographic maps qualitatively indicate that the LOS of the CS happens to be underdense almost all the way out to $z\approx0.3$. In other words, the CS region is avoided by prominent and rich superclusters, but is not necessarily the deepest underdensity in the maps. 

This claim disagrees with the conclusions by \cite{SzapudiEtAl2014} who found that there is no underdensity in the CS region at $z< 0.1$. Again qualitatively, we now see that the limited PS1 photo-z coverage due to the survey boundary at $Dec=-28^{\circ}$, marked by the black dashed box in the tomographic images in Figure 5, is inappropriate for detecting extended underdensities because there is no contrast with respect to the background they use for estimating the mean galaxy density. Moreover, a large fraction of the Eridanus supervoid located outside of the $\theta<5^{\circ}$ CS centre was not observed in the limited PS1 survey window.

\subsection{Line-of-sight density profile}

Following \cite{SzapudiEtAl2014}, we continue with the measurement of mean matter densities $\delta_{m} = \delta_{g}/b_{g}$ (corrected for deterministic linear galaxy bias) at different distances in the direction of the CS in a disk of $\theta=5^{\circ}$ around the centre. 

For completeness, we measure the angular density profile around the CS centre in the {\it full projected} 2MPZ galaxy catalogue. We show our results in the inset of Figure~6. The counts are consistent with a deeper core with $\delta_0 \approx -0.25$ at $\theta<5^{\circ}$ (aligned with the coldest part of the CS), while the mean density is reached at $\theta\approx25^{\circ}$ corresponding to $r_{0}^{\perp}\approx 200$ Mpc/h size. \cite{FinelliEtal2014} reported similar findings based on the projected WISE-2MASS galaxy catalogue. The tomographic imaging presented in Figure 5 also indicates that the central $\theta<5^{\circ}$ part of the CS region is typically surrounded by overdensities inside a more extended underdense area.

In Figure~6, we show both 6dF and 2MPZ results at redshifts $z<0.08$ where the constraining power of the more accurate 6dF spec-z data is adequate. The right side of Fig.~6 shows that the deepest part of the supervoid is located close to $z_{0}\approx0.14$, thus closer to us than reported by \cite{SzapudiEtAl2014}. More importantly, the supervoid appears to extend towards the lowest redshifts, again in disagreement with \cite{SzapudiEtAl2014}. We interpret these differences as an imperfection in the PS1 photo-z analysis due to small survey area, and the worse sampling in the WISE-2MASS-PS1 catalogue at the lowest redshifts. Note, however, that the PS1 measurements (data points and error bars taken from Figure 4 in \cite{SzapudiEtAl2014}) are consistent with 2MPZ within 1$\sigma$ errors.

The ability of resolving relatively small-scale structures with 6dF spec-zs and the high-quality photo-zs from 2MPZ allow us to notice the following details in our analysis, in agreement with the tomography:
\begin{itemize}
\item the CS lies close to the Fornax cluster region ($z<0.02$).
\item the Horologium supercluster's filament in its outskirts results in a short overdense part along the LOS.
\item there is an extra deepening in the 6dF counts compared to 2MPZ at $0.02<z< 0.06$ with several known and catalogued voids in between the two low-z over-densities Fornax and Horologium. 
\item the last 6dF bin is in great agreement with the 2MPZ bin of similar size, meaning that the two catalogues with different biases trace similarly the underlying dark matter field.
\item similar coherence is observable for PS1 and 2MPZ for the less accurately sampled bins at $0.15<z<0.20$ and $0.20<z<0.30$.
\end{itemize}

These observations can be interpreted as a manifestation of the smearing effect of the photo-z uncertainties, as we briefly discussed in Section 1, with shallower void centre in photo-z data and undetectable narrow over-densities along the line-of-sight without spec-z mapping. Future spec-z follow-up surveys beyond the 6dF window might reveal that the Eridanus supervoid contains further substructures. Recently, \cite{Naidoo2015} discussed the models of the ISW imprint of multiple isolated voids in alignment with the CS, and analyzed how the actual anomaly may be reduced by removing the ISW contribution.

In Fig.~6, we show the supervoid model $\delta_0 = -0.25$, $r_{0}^{\parallel}=500$ Mpc/h together with the matter density fluctuations. We also illustrate the smearing effect of photo-z errors on the shape of this model void. The blue dotted line shows the model curve without a $\sigma_{z}\approx 0.015(1+z)$ Gaussian smearing applied to the void profile to model the properties of the 2MPZ catalogue. For completeness, we added more significant smearing effect with a larger Gaussian photo-z error $\sigma_{z}\approx 0.05(1+z)$ in order to illustrate the capabilities of the 2MPZ data set in our problem. We inferred that voids comparable to the Eridanus supervoid can be precisely studied even in the presence of photo-z uncertainties, but the quantitative details of the problem of photo-z errors on voids should be worked out in a later project \citep[see also e.g.][]{Sanchez2016}.

We then perform a $\chi^{2}$ analysis using the 2MPZ data points and the $0.3<z<0.5$ measurement by \cite{GranettEtal2010}. We assume a diagonal covariance matrix thus no correlation between the different photo-z bins which is a defendable approximation given the width of the bins and the relatively low photo-z uncertainties in 2MPZ. The $\chi^{2}$ plane shown in Fig.~7 is estimated for the parameter space explored in Fig.~3 (d.o.f. = 6+1-2). The same supervoid candidates, and contours for the $\Delta T_{\rm ISW}$ imprint of voids are shown on top of the consistency measure of the LSS data and LOS model profiles.

We find that previously proposed spherical voids are barely consistent with the region of lower $\chi^{2}$ values ($\chi_{\rm min}^{2}/(6+1-2)\approx1$). Note that spherical supervoids of $r_{0}>200$ Mpc/h and/or $\delta_0 < -0.25$ are already inconsistent with the projected radial galaxy density profile that we presented in the inset of Figure~6. Nevertheless, the combination of available constraints on transverse and line-of-sight void size indicate significant elongation in the LOS. The elongated model supervoid of parameters $\delta_0 = -0.25$, $r_{0}^{\perp}=195$ Mpc/h, and $r_{0}^{\parallel}=500$ Mpc/h is not just consistent with the LSS data but in fact favored with some degeneracy between $\delta_0$ and $r_{0}^{\parallel}$. The angular $\Delta T_{\rm ISW}$ profile of such a supervoid, however, is not similar to the CS pattern (see also Figure~2), thus the explanatory power of the LSS evidence for a significant supervoid appears to be low.

The complicated angular and LOS shape of the supervoid, that we have already seen in Fig.~5 and Fig.~6, require both more sophisticated modeling and more accurate measurements. However, the evidence that supports the idea of an elongated void and a corresponding significant $\Delta T_{\rm ISW}$ imprint cannot be ignored.

\begin{figure}
\begin{center}
\includegraphics[width=90mm]{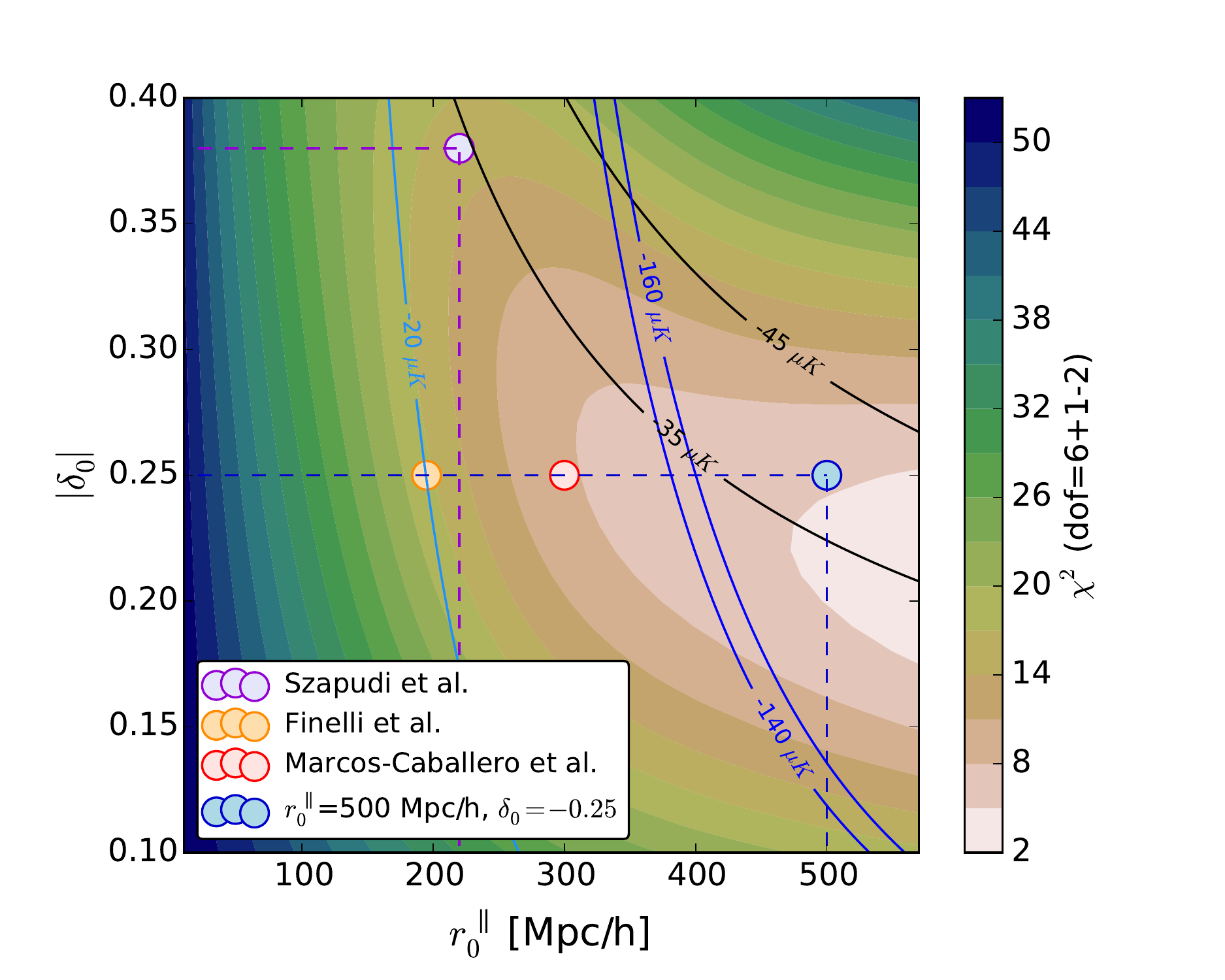}
\label{}
\caption{Consistency analysis of models and observations for the LOS density profile of the Eridanus supervoid. On top, we show the contours and symbols introduced in Fig.~3. The underlying colours and the corresponding colour bar illustrates the $\chi^{2}$ values considering different supervoid models and the combination of 2MPZ data and Granett et al. (2010) measurements, assuming a diagonal covariance matrix.}  
\end{center}
\end{figure}

\subsection{The Cold Spot antipode}

We extend the conventional LOS void profile analysis with the analysis of the antipodal direction. The motivation behind this analysis is the evidence in Fig.~6 that the Eridanus supervoid reaches our closest vicinity in the CS LOS, and there might be a continuation to the nearby antipodal regions. In fact, we find that this antipodal LOS traverses the Northern Local Supervoid, i.e. a prominent underdensity in the very local Universe studied for instance by~\cite{Lindner1997}.

After passing through the NLV, the antipodal LOS becomes overdense while approaching the rich Hercules supercluster close in the Coma Great Wall, and the very rich, so-called multi-spider supercluster Corona Borealis, i.e. an important member of the Sloan Great Wall. Another example of a multi-spider supercluster is Centaurus \citep{Courtois2013}. \cite{Einasto2011} found that the CBS happens to be in the intersection of the SGW and another large chain of superclusters, and it is a member of a huge system of rich superclusters, called the Dominant Supercluster Plane, located at right angles with respect to the Local supercluster \citep{Einasto1997}. 

These overdensities almost certainly mark the edge of the Eridanus supervoid that apparently extends to our location in space. Note that the far edge of the supervoid is poorly constrained because the 2MPZ and WISE-2MASS-PS1 surveys run out of galaxies at redshifts $z>0.2$, but DES will presumably map this range accurately in the future.

In Fig.~6, we show that the large-scale overdensity field, smeared by 2MPZ photo-zs, is consistent with $r_{0}^{\parallel}$=200 Mpc/h and $\delta_{0}=0.40$ in our framework. The 2MPZ galaxy counts at the CS are consistent with model supervoids shallower and longer than $\delta_{0}=-0.25$ with $r_{0}^{\parallel}$=500 Mpc/h (see Fig.~7), mostly because of the lack of constraining power at high distances from the centre of the supervoid. Note that a supervoid of radius $r_{0}^{\parallel}$=500 Mpc/h centred at $z_{0}=0.14$ already extends fully to our position in space, therefore additional constraints on this LOS density profile might come from local density estimates. Moreover, a combined modeling of the antipodal counts from 2MPZ, and presumably more precise future CS density constraints from DES and other surveys could be helpful for shrinking the allowed parts of this parameter space.

Our Fig.~8 is an updated version of a figure by \cite{Lindner1997}, showing the Dominant Supercluster Plane that happens to include the CS LOS that we are studying. We also added the approximate direction and location of the Draco supervoid, and named several prominent local supervoids and superclusters in the map, including Laniakea \citep{Tully2014}, the NLV, and the Southern Local Supervoid (SLV) \citep{Einasto1994}.

We can confirm our previous findings and make further observations. The CS line-of-sight:
\begin{enumerate}
\item avoids the local Laniakea supercluster
\item traverses both the NLV and SLV in the two antipodal directions (see also Fig.~6)
\item passes by the Horologium(-Reticulum) supercluster in the South (see also Fig.~5 and Fig.~6)
\item encounters the region of the Great Walls and the rich Hercules and Corona Borealis superclusters
\end{enumerate}

The Draco supervoid, defined by the projected WISE-2MASS galaxy data in~\cite{FinelliEtal2014}, may be similar to the NLV, or at least has a significant contribution for it to be a wide underdensity in the Northern sky, with presumably a rather oblate shape in the LOS.

The question arises: is this LOS that connects the most under- and overdense neighborhoods special? We create a full antipodal map for the smoothed \texttt{HEALPIX} maps {\it Planck}, 2MPZ, and the reconstructed and impainted 2MASS ISW \citep{rassat2013}. The latter attempts to predict the ISW signal based on the 2MASS density field, which is very similar to that of 2MPZ. Then we take the difference of the original and antipodal maps, as shown in Fig.~9 map-by-map, and observe the following features:

\begin{itemize}
\item the direction of the CS (yellow circle) is independently selected as the most dissimilar region in each map
\item the centre of the large-angle underdensity around the CS location is somewhat misaligned with respect to the actual CS centre, but the antipodal difference maps shows its extrema extraordinarily close to the nominal CS centre
\item the extended underdensity labeled as Draco supervoid by \cite{FinelliEtal2014} is the neighboring super-structure of the Great Walls region from another side than the Eridanus supervoid
\item the Draco supervoid in fact consists of several known voids, indicating that this large-angle underdensity is relatively close to us thus it presumably has an oblate shape
\item the direction of the Corona Borealis supercluster is surrounded by a set of other known voids
\end{itemize}

\begin{figure}
\begin{center}
\includegraphics[width=90mm]{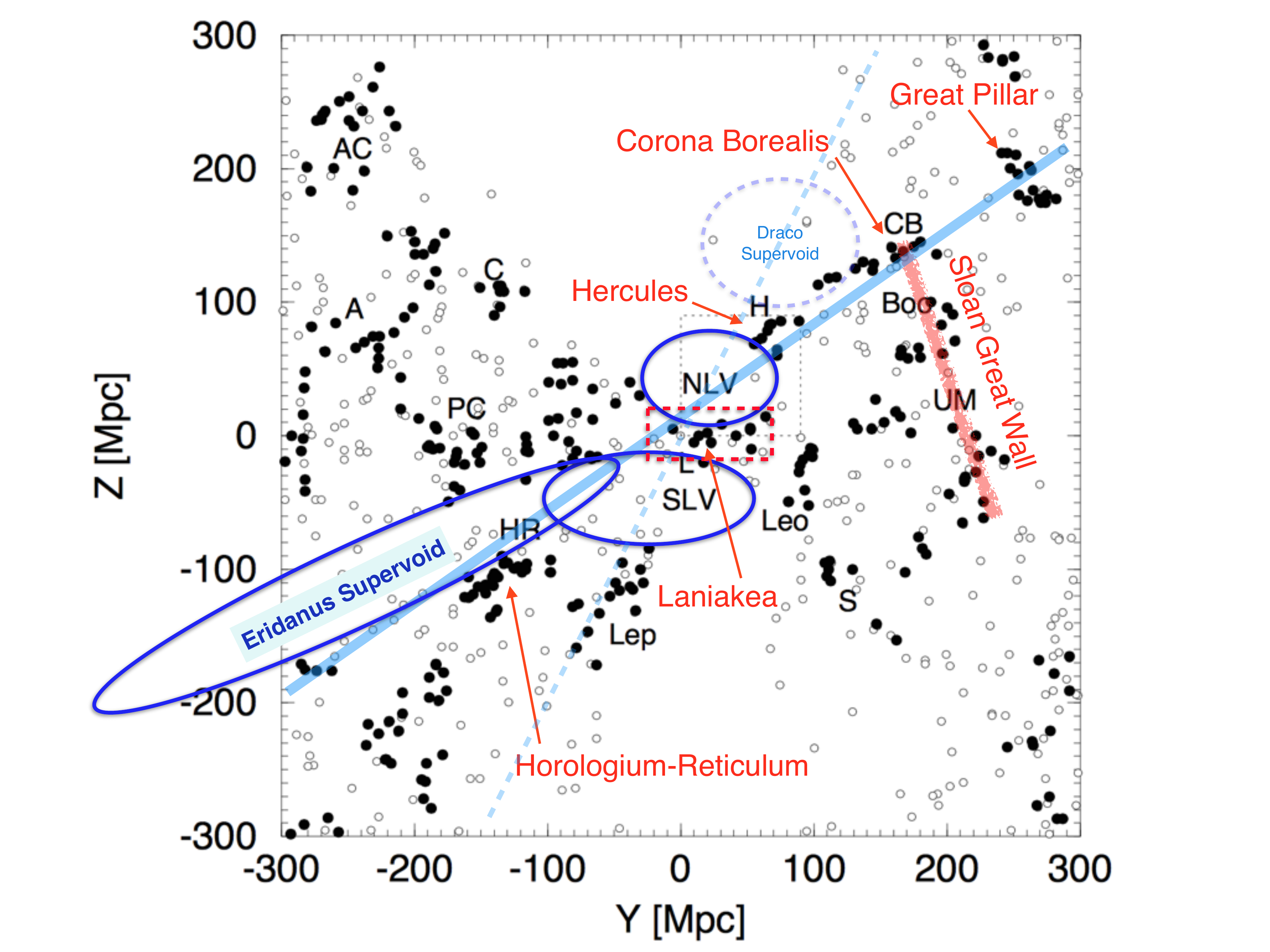}
\label{Lindner}
\caption{A modified version of Fig.~1 in Lindner et al. (1997). Filled black points mark rich clusters among Abell-ACO clusters (open circles) located in the Dominant Supercluster Plane of 200 Mpc/h thickness. The pale blue (dashed) band roughly indicates the line-of-sight that traverses the CS (Draco) centre and its antipode.}
\end{center}
\end{figure}

\begin{figure*}
\begin{center}
\includegraphics[width=58mm]{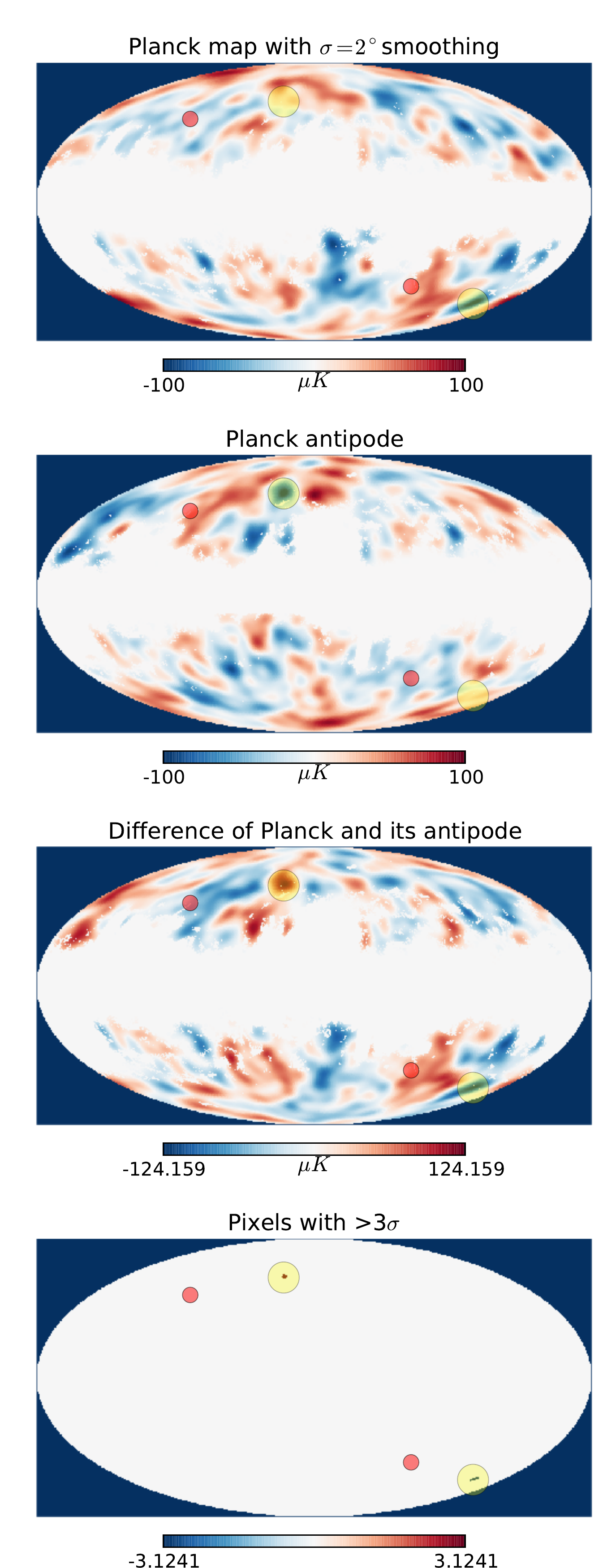}
\includegraphics[width=58mm]{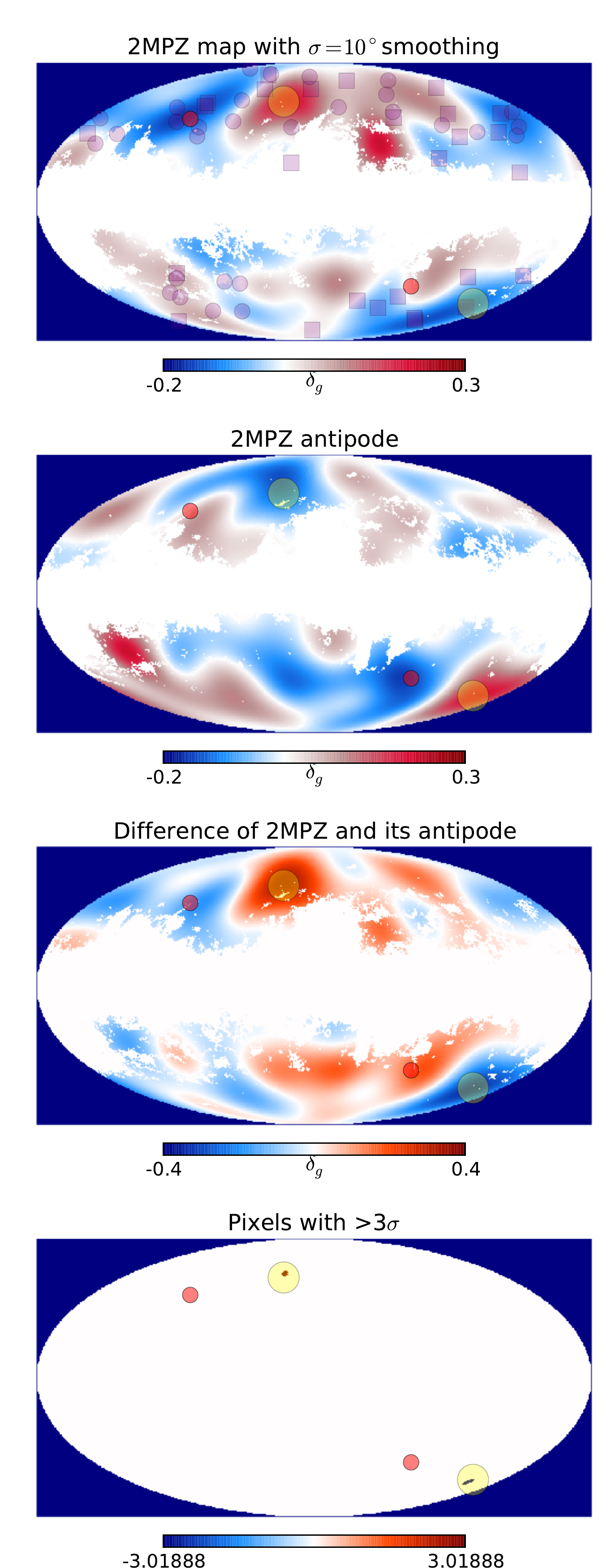}
\includegraphics[width=58mm]{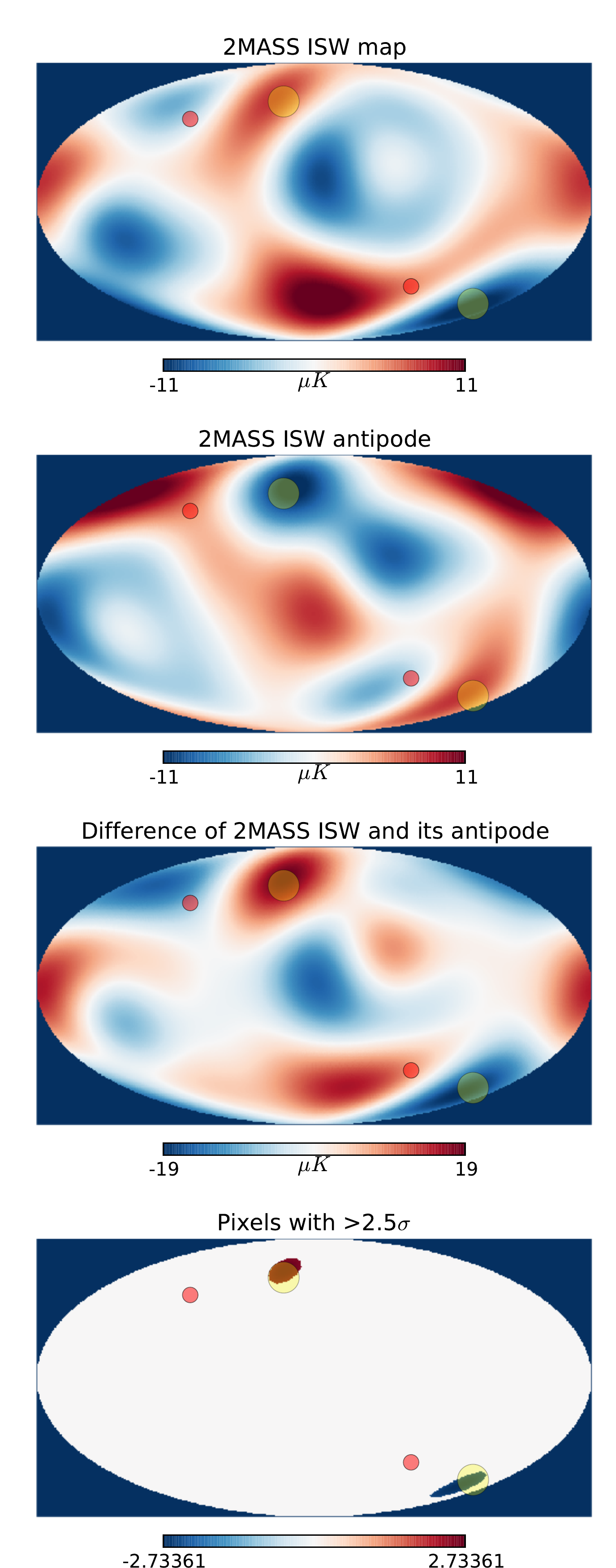}
\label{antipodes}
\caption{Analyses of antipodal differences in the Planck temperature map (left), the full 2MPZ density map (centre), and the reconstructed 2MASS ISW map by Rassat et al. (right). The direction of the CS (yellow circle) is independently selected as the most dissimilar region in each map. The approximate centre of the Northern Draco supervoid and its antipode are marked by the red spots. {\it Top-centre:} note that the direction of the Draco supervoid is abundant in voids defined by galaxy clusters identified in EEDTA (purple squares) and $B\&B$ (purple circles). The wider CS region also contains several voids from these void catalogues.}
\end{center}
\end{figure*}

The smoothing scales we use in this analysis were chosen {\it a posteriori} based on the characteristic angular sizes of the CS and the Eridanus supervoid. The CS represent an extreme event in the CMB at $\theta\approx5^{\circ}$ scales, while the total angular size of the Eridanus supervoid is $\theta \approx 25^{\circ}$. 

We performed tests using different filter sizes. Choices of Gaussian filters of roughly the half of these characteristic sizes result in a single $\sim3\sigma$ event in both maps. Slightly smaller filters yield 1-2 additional groups of pixels with $\sim3\sigma$ significance, but typically at smaller angular extent. We note that $\sigma=2^{\circ}$ or $\sigma=5^{\circ}$ filterings for the 2MPZ map, and considerations of additional $\sim2.5\sigma$ pixels in the $\sigma=10^{\circ}$ filtered map all select the region of the Shapley supercluster and its antipode as another (less significant) dissimilar region in the local galaxy distribution, but without significant counterpart in the antipodal CMB difference maps.

We then analyzed Gaussian CMB and LSS simulation maps, and found that such a local dissimilarity at the $\sim3\sigma$ level in the antipodal difference maps is not unusual on its own. The probability of having the most dissimilar direction in the CMB and in the LSS aligned within $\theta\leq5^{\circ}$ level appears to be small ($p_{\rm match}\sim10^{-4}$), but this depends on the choice of the smoothing scale and we have no {\it a priori} motivation for any smoothing level. Advanced probabilistic analyses would require prior probabilities that naturally disfavor the existence of large voids and strong ISW imprints based on the $\Lambda$CDM model predictions. Additionally, such advanced analyses should account for {\it look-elsewhere effects} associated with the smoothing analysis technique.

\section{Discussion and Conclusions}
We critically revisited the available observational constraints and theoretical considerations for unveiling the physical connection between the CMB Cold Spot and the Eridanus supervoid. 

The main motivations for such a follow-up analysis were the possible shortcomings of the density and size estimates for the supervoid based on the PS1 photo-z maps by \cite{SzapudiEtAl2014}, and the detection of significant LOS elongation of SDSS supervoids that also appear to leave anomalous imprint in the CMB \citep{Granett2015}.

We performed tomographic imaging in several redshift slices, and measured the density profile of the Eridanus supervoid along the line-of-sight using the 2MPZ photo-z data, supplemented by spec-z information from the 6dF survey at $z<0.1$. Our analysis provides important evidence for a LOS extent larger than the $r_{0}^{\parallel}\approx200$ Mpc/h size estimated previously. We detect potentially significant substructure in the form of the underdensities at the lowest redshifts, where \cite{SzapudiEtAl2014} found no depletion in their photo-z data. Our LSS data is consistent with an elongated supervoid of central depth $\delta_0 = -0.25$, inconstant transverse radius with projected size $r_0^{\perp}\approx 200$ Mpc/h (affected by extra deepenings in the central $r_0^{\perp}\approx 50$ Mpc/h or $\theta \approx 6^{\circ}$ region), and LOS size $r_0^{\parallel} \approx500$ Mpc/h.

Guided by the finding that the Eridanus supervoid appears to continue towards the lowest redshifts, we extended our LOS profile analysis into the antipodal direction, i.e. mapped the neighboring parts of the local Universe in the opposite direction. We discovered that the CS antipode crosses the Northern Local Supervoid, the rich Hercules supercluster, and the very rich Corona Borealis supercluster, effectively marking the edge of the supervoid. Therefore, this line-of-sight intersects the largest and richest supercluster systems in the local Universe, the Coma Great Wall and the Sloan Great Wall.

We then analyzed the full 2MPZ, {\it Planck}, and reconstructed 2MASS ISW data sets in order to test if this direction is significant and special in the observable sky. We found, remarkably, that the CS location is independently selected as the most dissimilar region in each map, pointing towards a physical connection between the CS and the Eridanus supervoid. The exact significance of this finding, however, slightly depends on the {\it a posteriori} selected smoothing scale that we apply to the maps.

Our ellipsoidal model for the ISW imprints of supervoids predicts significant contributions to the Cold Spot temperature at the $\Delta T_{\rm ISW} \approx -40 \mu K$ level. However, this rather strong imprint is not sufficient for explaining the central $\Delta T \approx -150 \mu K$ temperature decrement in the CS centre, and the predicted and observed angular profiles are different. Similar conclusions were reported by \cite{rassat2013} based on a reconstruction of the ISW map from the 2MASS survey data, by \cite{MarcosCaballero2015} who studied the ISW imprint of less extremely elongated supervoids, and by \cite{Naidoo2015} who estimated the effects of multiple voids in the CS direction.

The most important question besides the expected ISW imprint is the rarity of such an elongated supervoid in the $\Lambda$CDM framework. In general, similarly large voids of effective radius $R_{\rm eff}\geq 200$ Mpc/h (and central densities deeper than $\delta_0 = -0.25$) have been identified in SDSS-BOSS data sets\footnote{http://research.hip.fi/user/nadathur/download/dr7catalogue/} \citep{Nadathur2015}, or with photo-z tracers \citep{GranettEtal2008}. However, little is known about the actual shape of the largest voids, because the \texttt{ZOBOV} algorithm assigns an effective radius $R_{\rm eff}$ of a sphere for all underdensities of arbitrary shape. Modification of void finding algorithms and better understanding of voids identified in photometric data sets might reveal further details about this problem, and realistic prior knowledge on the expectation of atypical supervoids can be added to similar analyses.

On the theoretical side, spherical supervoids of $R_{\rm eff}\geq 300$ Mpc/h and $\delta_0 \approx -0.25$ are not expected to practically exist at $z<0.5$, assuming $\Lambda$CDM physics \citep{Nadathur2014}. However, there is firm observational evidence and theoretical expectation for $\mathcal{O}(10)$ supervoids of $R_{\rm eff}\approx 200$ Mpc/h and $\delta_0 \leq -0.25$ in the local Universe, and the currently unexplored nature of the interconnection of these already extended underdensities might provide surprises. 

The elongated shape that we reconstructed for the Eridanus supervoid is a property that makes the actual predictions more complicated. The same applies to large and relatively rare filamentary superclusters like the Sloan Great Wall. Whether Great Walls and elongated supervoids are more abundant than expected, has to be further investigated in large N-body simulations. Connections to other known local underdensities and anomalies should also be studied \citep{Schwarz2015,Boehringer2015,Whitbourn2014,Vikram2015,PadillaTorres2009,FloresCacho2009,KeenanEtal2012,Horvath2015}.

In any case, the more precise mapping we carried out for the Eridanus supervoid further enhanced its dimensions and pointed out connections to significant local voids and the Great Wall region. Given the size of these matter fluctuations, the analysis and revision of the Copernican principle is of high importance, especially in a statistical fashion \citep{Nadathur2013,Alonso2015}. In case more precise future mappings, and more sophisticated modeling will accurately elucidate the physical connection between the Cold Spot, the Eridanus supervoid, and the region of the Great Walls, the Eridanus constellation may become the most promising laboratory for Dark Energy experiments.

\section*{Acknowledgments}

We thank Seshadri Nadathur, Shaun Hotchkiss, David Alonso, Airam Marcos-Caballero, Fabio Finelli, Francesco Paci, Istv\'an Szapudi, and the anonymous referee for useful comments that improved the paper. We also thank Krishna Naidoo, Ofer Lahav, and Aurelien Benoit-L\'evy for informative discussions. AK thanks M\'arton Vargyas, and JGB the Theory Division at CERN, for their hospitality, where the main ideas of the project were discussed. Funding for this project was provided by the Spanish Ministerio de Econom\'ia y Competitividad (MINECO) under projects FPA2012-39684, FPA2013-47986, and Centro de Excelencia Severo Ochoa SEV-2012-0234 and SEV-2012-0249.

\bibliographystyle{mn2e}
\bibliography{refs}
\end{document}